# Enhancement of $CO_2$ conversion by counterflow gas quenching of the post-discharge region in microwave plasma sustained by gyrotron radiation

N.V. Chekmarev[*], D.A. Mansfeld, A.V. Vodopyanov, S.V. Sintsov, E.I. Preobrazhensky, M.A. Remez

Institute of Applied Physics, Russian Academy of Sciences, Nizhny Novgorod, Russia

[*]Corresponding author: N.V. Chekmarev, chekmarev.nikita7@gmail.com, 46 Ul'yanov Street, Nizhny Novgorod, 603950, Russia

Abstract

A threefold increase in the $CO_2$ conversion and energy efficiency due to the cooling of the post-discharge region by the counter gas flow has been achieved in the plasma of an atmospheric pressure discharge supported by microwave radiation of a gyrotron with a frequency of 24 GHz in a carbon dioxide gas flow. The role of convective heat transfer in the process of gas mixture cooling in the post-discharge region has been experimentally demonstrated. At nitrogen quench gas flow of 4.5 l/min, the $CO_2$ conversion was 23.8 % and energy efficiency was 19.7 %. The possibility of using the flow of cooled gas mixture ($CO_2, CO, O_2$) taken from the reactor as quenching gas has been experimentally demonstrated, which made it possible to achieve a $CO_2$ conversion degree of 23.4 % and to eliminate the problem of dilution of reaction products by third-party gases. Based on numerical modeling, it is shown that the increase in the conversion degree upon the destruction of the plasma torch structure is due to the increase in heat exchange with the surrounding atmosphere, and the efficiency of this destruction is determined by the velocity and density of quenching gas.

1. Introduction

The growth of greenhouse gas emissions into the Earth's atmosphere and, therefore, the increase in global temperature by several degrees, is alarming the entire world community. At present, enormous efforts are aimed at finding methods to capture, process and store (CCS and CCU) carbon dioxide as the main product of fossil fuel combustion [1], [2]. The most preferable from the economic point of view is the conversion of carbon dioxide into marginal chemical products using energy from renewable sources. In particular, the decomposition of carbon dioxide produces carbon monoxide, which can serve as a feedstock for the Fischer-Tropsch process and for the production of hydrocarbons, including methanol, dimethyl ether, and others [3,4]. Among the most attractive approaches are plasma chemical methods for carbon dioxide conversion in the plasma of various atmospheric pressure discharges, where dissociation of $CO_2$ molecules occurs due to stepwise excitation of vibrational degrees of freedom by electrons and excited $CO_2$ molecules [5]. From the entire variety of atmospheric pressure discharges studied over the last decades, the dielectric barrier discharge (DBD), gliding arc (GA) discharge, and microwave discharge have proved to be the most promising for achieving maximum degrees of conversion and energy efficiency [6].

In cold (nonequilibrium) DBD plasma, it is possible to achieve maximum degrees of $CO_2$ conversion only at high specific energy input, which does not allow simultaneously to achieve high energy efficiency. In the warm plasma of gliding arc discharge it is possible to achieve conversion of about 10% with energy efficiency of about 30% [6]. However, attempts to achieve a significant increase in the $CO_2$ conversion by modifying the discharge area and electrode configuration, optimization of gas flows, do not bring the desired result [6]. Microwave discharge, being in principle electrodeless, seems to be the most probable candidate for realization of a pilot prototype of plasma-chemical installations both for $CO_2$ decomposition processes [7,8] and, for example, for nitrogen fixation processes [9]. Microwave discharge is most effective at low pressures, allowing to achieve high degrees of carbon dioxide conversion, but this mode of operation requires the use of expensive vacuum equipment, which makes it difficult to scale up to the industrial level. With pressure increase to atmospheric value, the gas temperature rises to values of about 6000 K and $CO_2$ conversion drops sharply as a result of V-T relaxation down to values of about 10%. Nevertheless, the process of $CO_2$ to $CO$ conversion can achieve appreciable energy efficiency approaching the thermodynamic limit of 50% [10,11]. Thus, as recent studies show, in the warm plasma of microwave and GA discharges at atmospheric pressure it is possible to achieve approximately the same values of conversion and energy efficiency of the $CO_2$ decomposition process [6,12].

One of the factors reducing the efficiency of plasma-chemical methods of $CO_2$ decomposition in warm plasma is the occurrence of backward recombination reactions of $CO$ in $CO_2$ in plasma ($CO + O + M \rightarrow CO_2 + M$), ($CO + O_2 \rightarrow CO_2 + O$), the role of which becomes particularly significant at temperatures of 2000-3000 K [13]. Most authors agree that the most effective way to suppress recombination reactions is rapid ($10^6 - 10^7\ K/s$) cooling (quenching) of the gas mixture, which leads to "freezing" of the state with maximum conversion (ideal quenching) [14].

In addition, lowering the temperature can lead to further dissociation of excited $CO_2$ molecules as a result of the theoretically predicted "super ideal quenching" effect ( $CO_2(v) + O = CO + O_2$ ) [14]. Various approaches have been proposed to organize the quenching process, including operation in pulsed heating mode [15,16], introduction of water-cooled rods into the plasma region [17], but a really significant quenching effect was experimentally demonstrated by Hecimovic [18] in microwave discharge at a pressure of 900 mbar, when it was possible to increase the $CO_2$ conversion from 5% to 35% using a water-cooled nozzle in the post-plasma region. The authors attributed the increase in conversion to the mixing of the hot plasma gas and the surrounding cooler gas, which was later confirmed by simulation results, which, in particular, showed [19] that the presence of the nozzle leads to better convective cooling by forcing the cold gas near the walls to mix with the hot gas in the center of the reactor, and the water-cooled nozzle walls also contribute to better gas cooling. The effect of quenching on suppressing recombination reactions has also been shown to be particularly significant at low flow rates when recombination is the most limiting factor in the conversion process. In the work of Mercer et al. [20] demonstrated both experimentally and using CFD modeling the significant effect of a nozzle in the plasma of a microwave discharge on changing the characteristics of the gas flow, facilitating the rapid transport of the generated $CO$ to the reactor outlet (instead of participating in recombination reactions).

In this study, it is proposed to use a cold gas stream injected into the post-discharge region of the microwave discharge plasma for quenching. Cooling of the plasma torch in this case is achieved due to two effects: effective heat exchange with the surrounding atmosphere and heat exchange directly with the injected gas. An important feature is the research in the previously developed waveguide reactor (plasmatron) [21], in which the plasma is created and maintained by continuous microwave radiation of a technological gyrotron with a frequency of 24 GHz and a power of $20 - 5000 W$. The use of shorter-wavelength radiation compared to the traditionally used $2.45\ GHz$ microwave discharges allows to significantly increase the specific energy input and plasma density, which in turn can lead to an increase in the rate of plasma-chemical reactions [22]. In particular, in our previous work, it was shown that the use of focused gyrotron radiation with a frequency of 24 GHz made it possible to maintain the discharge at atmospheric pressure under nonequilibrium conditions and to achieve significant degrees of conversion both in the flow of pure $CO_2$ and in the mixture with argon [23].

2. *Experimental setup*

All experiments were carried out in the plasma of microwave discharge at atmospheric pressure at the installation, the scheme of which is presented in Fig. 1. Microwave heating is provided by continuous electromagnetic radiation of a technological gyrotron (frequency 24 $GHz$, $H_{11}$ mode, linear polarization), which enters the reactor through a water-cooled vacuum window made of boron nitride. The reactor is an extension of the supersized waveguide path of the gyrotron with an inner diameter of 32.6 mm. In order to increase the radiation power density, a copper truncated cone with a length of 60 mm and an exit hole diameter of 10 mm is attached to the waveguide path. The radiation power in the experiments described in the paper did not exceed 600 W, which corresponds to the field amplitude in the reactor output section of 900 V/cm. According to the results of calorimetric measurements, the absorption coefficient of microwave radiation in the plasma was 60-70%.

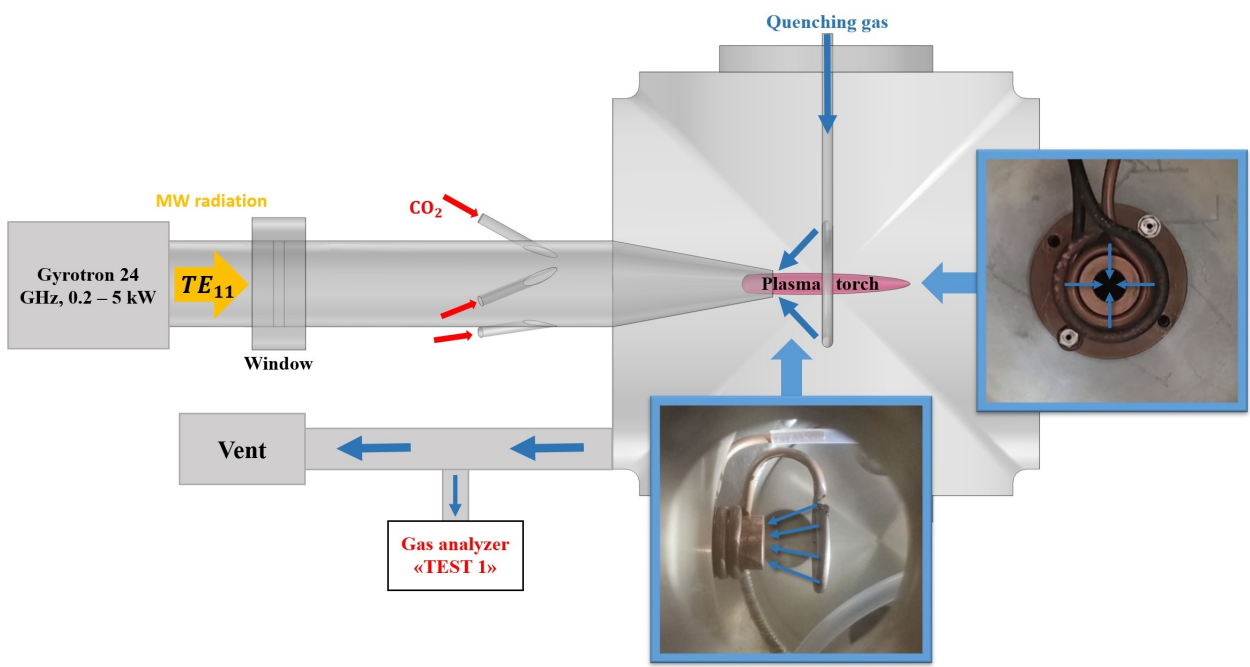

Figure 1. Schematic diagram of a waveguide plasmatron with combined gas supply.

The supply of carbon dioxide to the plasmatron is carried out through three symmetrically located tubes, making an angle of 30° with the plasmatron cylinder. The tangential inlet allows the formation of swirling gas streams, thus limiting the area of contact of hot plasma with the walls of the waveguide. The plasmatron is connected via a standard CF160 vacuum connection to an expansion chamber with a volume of about 10 liters, designed as a standard six-pass cross. The quenching gas supply in the post-discharge region is carried by a copper tube (internal diameter 2 mm) placed in the expansion chamber opposite the reactor outlet through four symmetrically arranged 1 mm diameter holes. All gases are supplied to the plasmatron at atmospheric pressure using Bronkhorst gas mass flow controllers in the range of 0-15 l/min (ANR) with an accuracy of 0.5%.

Part of the gas stream leaving the expansion chamber was sampled at a rate of 500 cc/min for subsequent analysis of reaction products in the gas analyzer "TEST-1", which contains oxygen ($O_2$), carbon monoxide ($CO$) and carbon dioxide ($CO_2$) sensors. The electrochemical $O_2$ cell in the measurement range of 0-21 vol.% has an absolute error of ±0.5 %. The wide-range optical $CO_2$ cell allows measurements in the range of 0-100 vol.%, but at high $CO_2$ concentrations its error is ±1%. The most accurate measurements are made by the $CO$ optical cell (Smartgas) in the 0-30% range with an error of ±0.3%. All sensors were calibrated over the entire operating range with calibration gas mixtures (relative accuracy <0.3%). Gas concentrations were measured after steady-state values were established, which ranged from 70 to 100 minutes depending on the gas flows.

The degree of $CO_2$ conversion, defined as the fraction of decomposed carbon dioxide molecules, was calculated by the formula:

$$K_{co} = \frac{\varphi_{co}(F_{solvent}+F_{co_2})}{F_{co_2}\left(1-\frac{1}{2}\varphi_{co}\right)} \cdot 100\%, \tag{1}$$

where $F_{CO_2}$ ($l/min$) is the carbon dioxide flow, $F_{solvent}$ ($l/min$) – is the flow of all gases supplied the chamber except $CO_2$, $\varphi_{co}$ – the measured $CO$ concentration in the chamber. In the paper, the diluent gas is understood as the quench gas flow in the post-discharge region, which practically does not enter the discharge region. In the case when quenching gas is absent formula (1) takes a usual form for determining the conversion degree in pure $CO_2$ [24].

$$K_{co} = \frac{\varphi_{co}}{\left(1-\frac{1}{2}*\varphi_{co}\right)} \cdot 100\%$$

It should be noted that both formulas automatically take into account the increase in gas volume during the $CO_2$ decomposition reaction. The formulas for determining the error in estimating the conversion and energy efficiency are given in the Supplementary file.

Energy efficiency, a parameter that determines how efficiently the decomposition process runs compared to the standard enthalpy of reaction, was determined by the formula [25]:

$$\eta(\%) = K_{co} \cdot \Delta H/SEI,$$

where $\Delta H = 2.93$ ($eV/molec$) – specific enthalpy of decomposition of $CO_2$ in $CO$ and $O_2$ at standard conditions, and specific energy input $SEI$ $\left(\frac{eV}{molec}\right)$ per one $CO_2$ molecule by the formula:

$$SEI\left(\frac{eV}{molec}\right) = \frac{P_{input}(W) \cdot 60(sec/min)}{F_{CO_2}(l/min)} \cdot 2.54 \cdot 10^{-4} \left(\frac{eV \cdot l}{J \cdot molec}\right),$$

where $P_{input}(W)$ – input microwave radiation power, $F_{CO_2}(l/min)$ – flow of $CO_2$.

### 3. Microwave discharge description

The discharge is initiated by short-term insertion of a thin metal wire into the cone. The discharge supported in the plasmatron by microwave radiation in the flow of $CO_2$ at atmospheric pressure consists of two qualitatively different regions. Inside the conical waveguide, the discharge is a plasma formation localized in the region of maximum electric field of the wave (near the exit) and oriented along the field direction. With increasing power, the discharge shifts deep inside the waveguide toward the microwave radiation. In the central near-axis region of the discharge, according to estimates from optical measurements of the rotational temperature of $CO_2$ molecules, the gas temperature can reach values of 6000-7000 K. In this zone, the chemical equilibrium is strongly shifted towards the reaction products and almost complete decomposition of carbon dioxide occurs: $2CO_2 \rightarrow 2CO + O_2$ (see Fig. 2a). The central region of direct reactions is streamlined by colder gas from the periphery of the waveguide, which passing through the strongly heated plasma forms a plasma torch with a length of 6 - 10 cm and a diameter of about 1 cm at the output of the plasmatron. Thus, a region of reverse reactions is formed, in which the reaction mixture is cooled mainly due to convective heat exchange with the atmosphere formed in the chamber, which leads to a shift of the chemical equilibrium towards the initial $CO_2$.

The cooling of the reaction mixture from 3000 K to 2000 K plays a key role, since it is in this temperature range that the rate of reverse reactions increases: $CO + O_2 \rightarrow CO_2 + O$; $CO + O + M \rightarrow CO_2 + M$. With sufficiently slow cooling, the gas mixture, passing sequentially through all equilibrium states, will completely change back to the initial $CO_2$, therefore, to realize the effective decomposition of $CO_2$, it is necessary to "fix" the composition of the gas mixture by its rapid cooling.

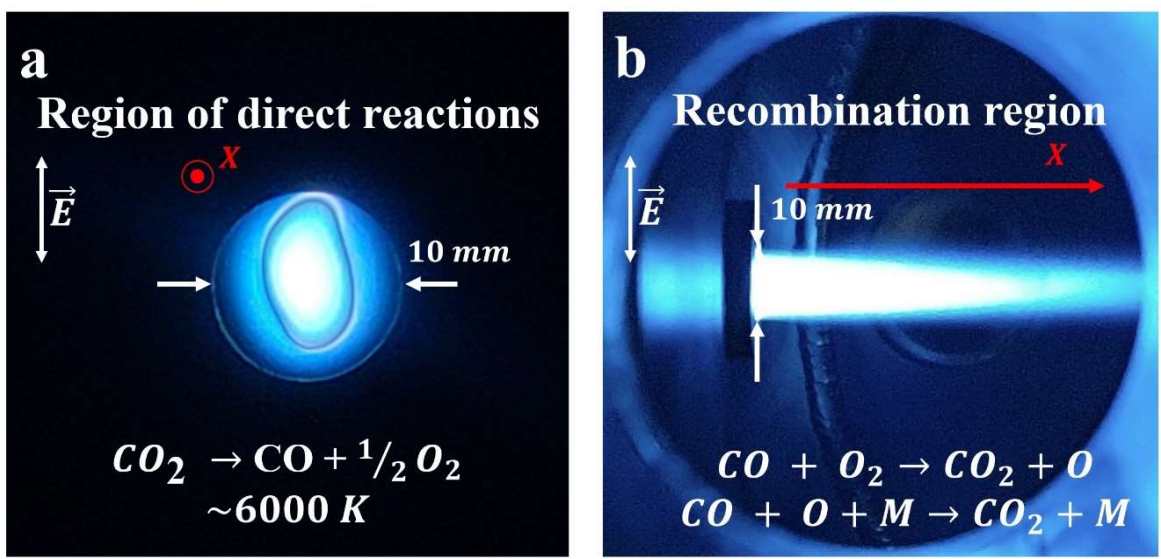

Figure 2: Photograph of the discharge cross section (a) and photograph of the plasma torch exiting the cone into the expansion chamber(b). $F_{CO_2} = 2,6\ l/min, P = 600\ W$. Inside the waveguide (a), a bright region elongated along the electric field direction is observed, streamlined by a cooler gas.

The plasma torch, after leaving the cone, cools rapidly, interacting with the atmosphere of the expansion chamber. To demonstrate the influence of convection heat exchange between the plasma torch and the chamber atmosphere on the process of $CO_2$ decomposition, an experiment was made to limit the post-discharge region. For this purpose, a copper tube with a length of 70 mm and an inner diameter of 20 mm was installed at the outlet of the conical section of the plasma torch (see Fig. 3).

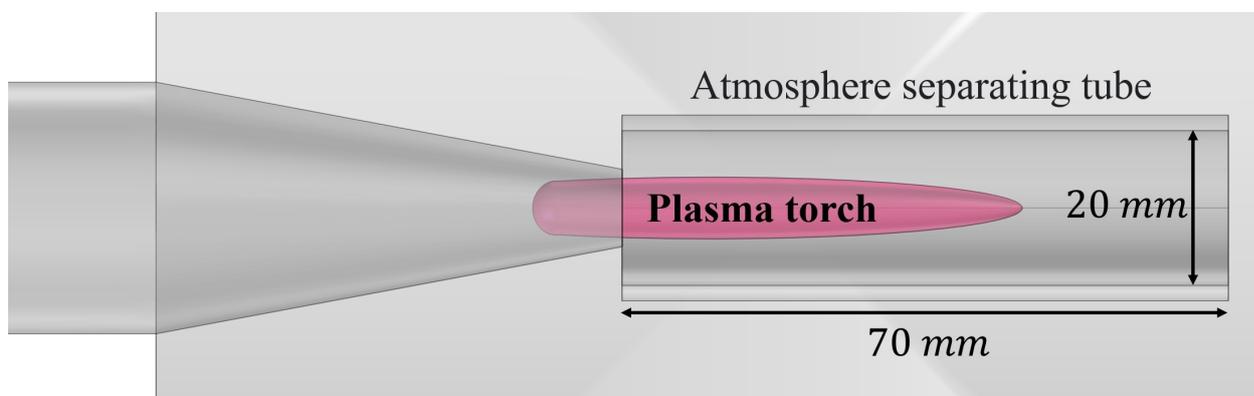

Figure 3: The scheme of the experiment to limit gas mixing in the post-discharge region.

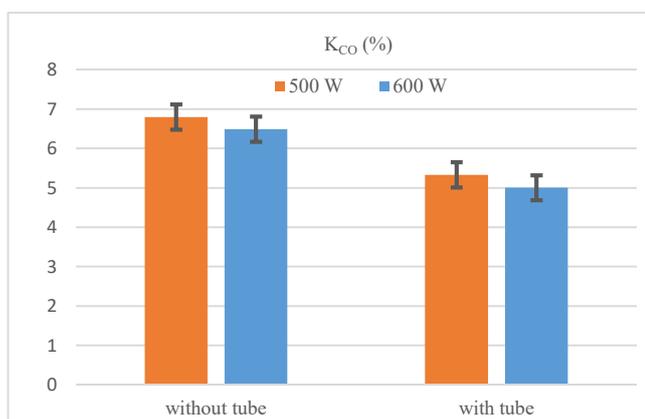

Figure 4: Effect of a tube placed on the output of the plasmatron on the conversion degree at carbon dioxide flow $F_{CO_2} = 2.6 \ l/min$ and at 600 W and 500 W of microwave heating power.

Figure 4 shows a comparison of conversion with and without a tube attached to the plasmatron outlet for two different values of heating power at a $CO_2$ flow of 2.6 l/min. It can be seen from Fig. 4 that the conversion decreases by more than 20% when the post-discharge region is bounded. Such a significant decrease in the conversion rate in the presence of a limiting tube additionally demonstrates the role of convective heat transfer in the cooling of the gas mixture. Also, in addition to heat transfer, the temperature decrease during the expansion of the gas leaving the waveguide opening, may also play a role. As can be seen from Fig. 4, the increase in heating power from 500 W to 600 W also does not give an increase in conversion. Increasing the power leads not only to an increase in the gas temperature in the near-axis zone of the discharge, but also to an increase in the length of the plasma torch in the post-discharge region, which also does not lead to an increase in the $CO_2$ conversion rate. This indirectly indicates that the increase in the plasma torch length negatively affects the conversion, since the process of destruction of $CO_2$ molecules predominantly occurs in a short time at high temperature inside the waveguide, and in the plasma torch there is a reverse recombination, and the longer this path, the more $CO$ molecules recombine back into $CO_2$.

4. *Quenching with a nitrogen.*

Achievement of the most effective quenching of reaction products and organization of the fastest heat exchange of the plasma torch with the surrounding atmosphere is possible due to injection of the quenching gas flow in the post-discharge region towards the main flow. The interaction of gas flows leads to changes in the shape of the plasma torch, including its complete "destruction", which, however, can lead to more intense heat exchange and gas temperature reduction. In the experiment, nitrogen $N_2$ was used as quenching gas, which was fed in the direction of the plasma torch coming out of the waveguide. Figure 5 shows photographs of the plasma torch in the post-discharge region taken at different values of the quench gas $F_{N_2}$ flow. The geometrical dimensions of the torch in the absence of quenching (Fig. 5a) and at quench gas flows less than 1L/min (Fig. 5b) are practically the same. However, with increasing nitrogen flow, there is a decrease in the length and an increase in the transverse size of the plasma jet (see Fig. 5 c,d,e). Starting from values of $F_{N_2}$ = 4 l/min, the plasma torch almost completely collapses and is near the exit of the cone (Fig. 6f). With further increase of quench gas flow, no visual changes are observed.

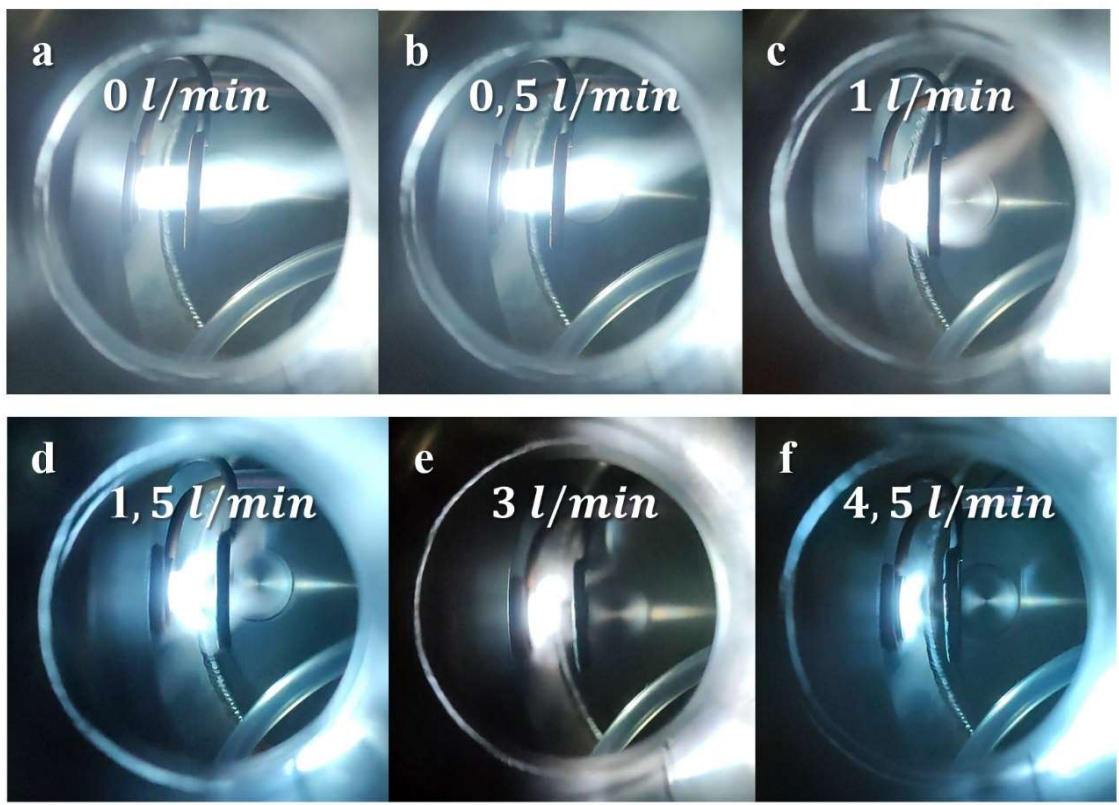

Fig.5 Photograph of plasma torch at different quench gas flows of $N_2$ from 0 to 4.5 l/min. $F_{CO_2}$ = 2.6 l/min, $P$ = 600 $W$. In the absence of quenching gas, the length of the plasma torch is 6-10 cm and the diameter is about 10 mm (a). At low quench gas flows up to 1 l/min, the shape and size of the plasma plume do not change significantly (b). With further increases in quench gas flow, the apparent diameter of the torch increases up to 20 mm and the length decreases to 10 - 15 mm (d). At high flows, the torch is almost completely destroyed, escaping from the plasmatron only by 5 mm (e, f).

Figs. 6 and 7 show the dependences of conversion degree and energy efficiency on quench gas flow $N_2$, at a microwave radiation power of 600 W and carbon dioxide flow $F_{CO_2}$ = 2,6 l/min. The conversion shown in Fig. 6 is calculated from the readings of the most accurate optical $CO$ cell.

This article was submitted to the Journal of Energy Chemistry 28.11.2023

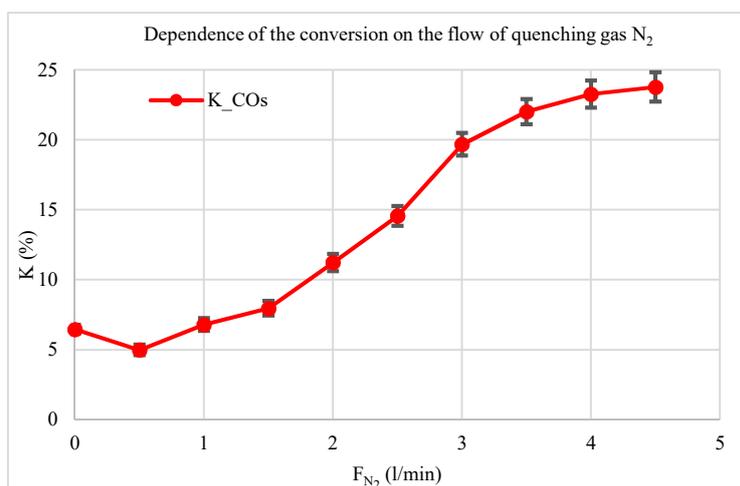

Figure 6. Dependence of $CO_2$ conversion on quench gas flow. Carbon dioxide flow $F_{CO_2} = 2.6 \, l/min$, microwave power $P = 600 \, W$

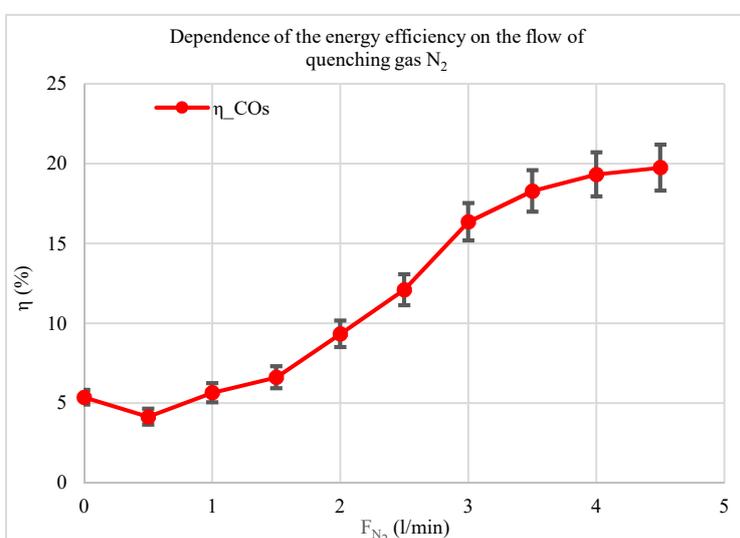

Figure 7. Dependence of energy efficiency on quench gas flow. Carbon dioxide flow $F_{CO_2} = 2.6 \, l/min$, microwave power $P = 600 \, W$

It can be seen from the graph in Fig. 6 that at a minimum flow of quenching gas, when the dynamic effect on the structure of the plasma torch is minimal, the effect on the degree of conversion is negligible. With an increase in the quenching gas flow, the conversion rate increases 3.5 times from 6.5% at $F_{N_2}=0$ (no quenching) to 23.8% at $F_{N_2} = 4.5$ l/min. Qualitatively, this dependence can be explained by the fact that with the growth of the quenching gas flow, the cooling rate in the post-discharge region also increases, and then the maximum quenching efficiency is achieved for a given power and a given quenching gas supply configuration. With a further increase in the nitrogen flow, the increase in the degree of conversion stops, and the discharge itself becomes unstable and extinguishes. With a fixed $CO_2$ flow and heating power, the specific energy supply also does not change ($SEI = 3.53 \, eV/molec$), therefore, the nature of the energy efficiency dependence (Fig.7) repeats the dependence graph in Fig. 6. The maximum energy efficiency η = 19.7% is also achieved with a quenching gas flow of 4.5 l/min.

To evaluate the possibility of formation of nitrogen oxides ($N_xO_y$) during the interaction of nitrogen flow with plasma containing oxygen atoms and molecules at temperatures of 2000-3000 K, at a flow of $F_{N_2}$ = 4.5 l/min, gas samples were taken for quantitative analysis using a chromatography-mass spectrometer Agilent 6890/MSD 5973N. As it follows from the analysis the concentrations of $NO$, $N_2O$ and $NO$ are extremely low - at the level of 10-3 - 10-4 vol.%., i.e. binding of oxygen atoms with nitrogen practically does not occur and the addition of nitrogen does not affect the accuracy of measurements.

5. *Quenching with a gas mixture taken from the reactor.*

Although the discharge in the above experiments is maintained in pure $CO_2$, the use of nitrogen as quench gas in industrial applications complicates the subsequent separation procedure of the output products, which contributes significantly to the economic feasibility calculation [6]. It is possible to use as quench gas the cooled mixture of reaction products - $CO, CO_2, O_2$, taken from the expansion chamber by a pump as shown in Figure 8. In this case, the gas balance in the reactor is not disturbed because exactly the same volume of gas is returned to the chamber as is withdrawn.

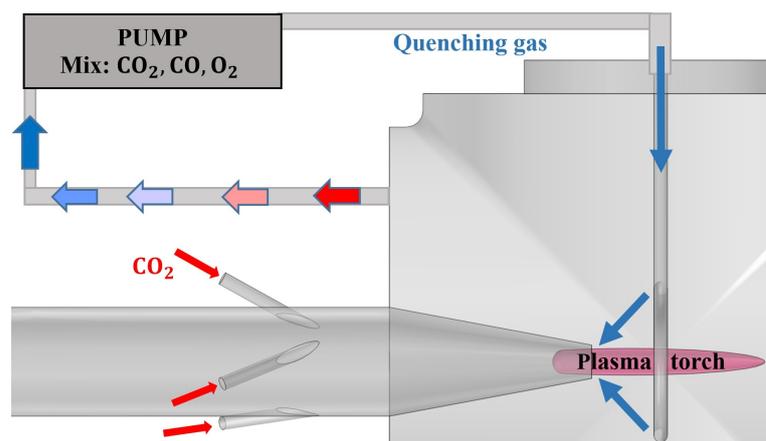

Figure 8. Scheme of quenching by flow of cooled gas mixture from the reactor.

Figure 9 shows a comparison of the carbon dioxide conversion with quenching by nitrogen flow and with quenching by gas mixture flow from the reactor. Quenching with the gas mixture from the reactor proved to be even more efficient than quenching with nitrogen, since the maximum conversion degree of 23.4% was achieved already at a quench gas mixture flow of $F_{quen}$ = 3 $l/min$. The maximum energy efficiency was 19.4%. However, further increasing the quench gas flow above 3 l/min leads to unstable torch combustion. Thus, the use of a gas mixture of $CO_2$ decomposition reaction products for "destruction" of the plasma torch allows to achieve a significant degree of conversion without additional costs for the use and subsequent separation of quench gas.

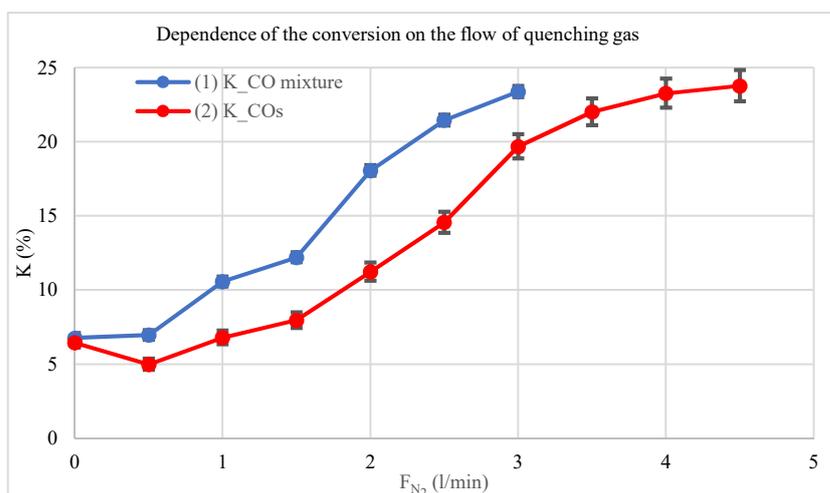

Figure 9. Comparison of quenching by nitrogen flow and flow of cooled gas mixture withdrawn from the chamber. $F_{CO_2} = 2,6 \ l/min$, $F_{que} = 3 \ l/min$, $P = 600 \ W$.

## 6. Discussion

We assume that we can distinguish two main mechanisms that qualitatively affect the efficiency of the quenching process of the plasma torch by the counter gas flow. First, it is the heat exchange between the plasma torch leaving the waveguide and quenching gas; the efficiency of such heat exchange is determined by the molar heat capacity of the quenching gas. Secondly, the dynamic effect of the quench gas flow on the torch structure. When dividing a single stream of hot gas from the plasmatron into many jets, heat exchange with the cooler atmosphere formed in the chamber is much faster. This dynamic effect is primarily determined by the transferred specific momentum, which depends on the density and velocity of the quenching gas. A comparison of the quenching efficiency of several different gases has been carried out previously in a microwave discharge in the flow of $Ar + CO_2$ mixture [26]. In particular, it was shown that for the same volume flow rate, molecular quenching gases with higher molar heat capacity are more effective in terms of quenching efficiency than single-atom gases, as well as gases with higher density were more efficient than less dense gases. Also, it was found [26] that the greatest influence on plasma torch breakup, and hence on quenching efficiency, is the quenching gas velocity, characterized by the flow rate. Apparently, a similar picture is observed in the experiment discussed in this paper. At a minimum quench gas flow, as one would expect, there is no effect on the degree of conversion because, first, the heat capacity of the gas flow is insufficient for effective heat transfer from the plasma torch, and second, the dynamic effect of quench gas jets is insufficient to destroy the structure of the torch and more effective mixing with the atmosphere inside the chamber. As the quench gas flow increases, the rate of cooling of reaction products in the post-discharge region also increases, which results in maximum quenching efficiency for a given power and a given configuration of quench gas channels.

For a more detailed understanding of the process of quenching by counterflow gas, numerical modeling of carbon dioxide decomposition in the discharge supported in the waveguide plasmatron by microwave radiation with the frequency of 24 GHz at atmospheric pressure was carried out. At the initial stage in the package COMSOL Multiphysics solved the problem of calculating the distribution of electromagnetic fields in the waveguide in order to determine the field strength region at which the conditions for discharge burning are realized. A detailed description of the problem is given in the Supplementary file. Figure 10a shows the distribution of the rms value of the electric field inside the conical part of the plasmatron and at its exit. The periodic structure in the left part is explained by the presence of reflected radiation, the power of which does not exceed 4% of the generated one. The zone with field strength (≥ 700 V/cm) with a characteristic size of ~ 8 mm, is located near the exit hole of the cone.

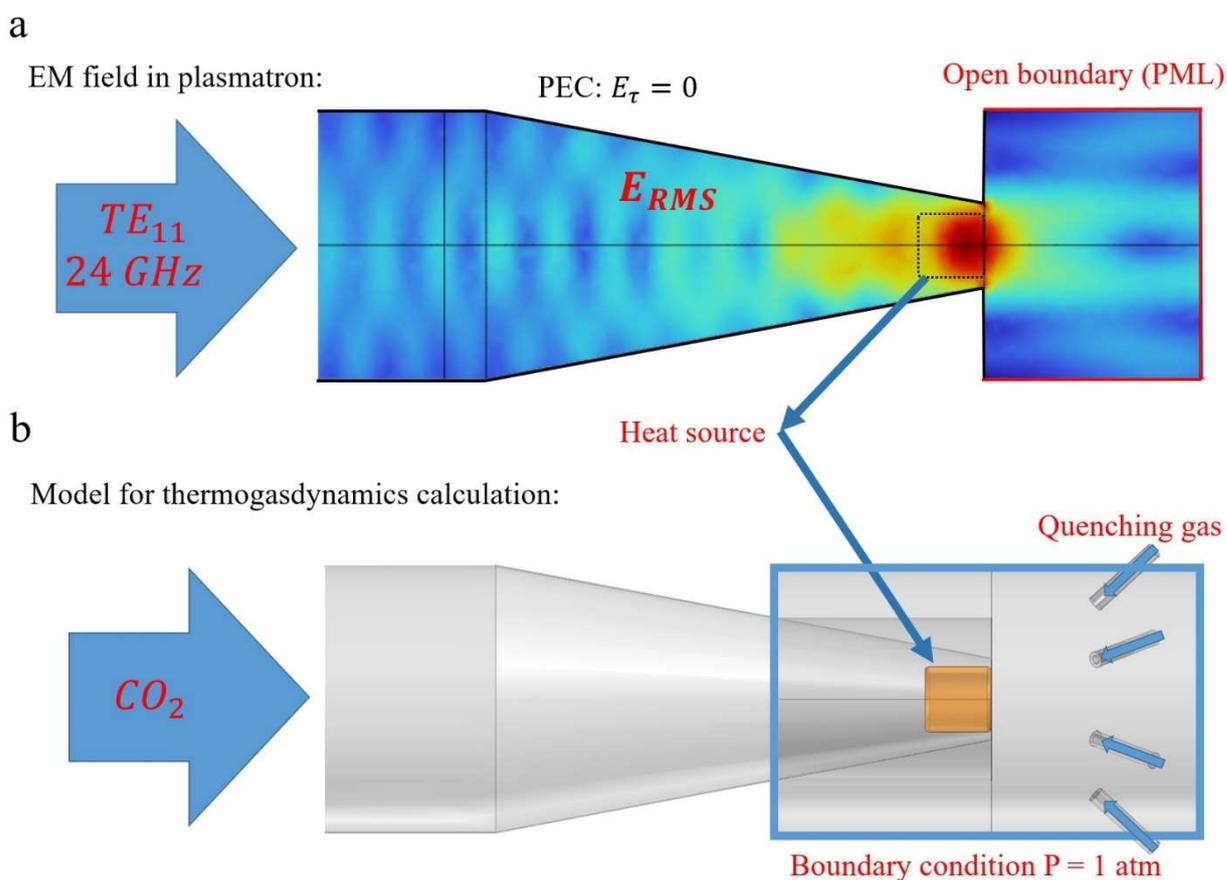

Figure 10. (a) Calculated in COMSOL RMS value of the electric field strength in the plasmatron and at its exit. The arrow indicates the region of the largest amplitude of the intensity. (b)Model of the plasmatron for calculating gas dynamics and heat transfer. The arrow indicates the heat release region, the position of which corresponds to the region of the electric field strength amplitude, which is sufficient to maintain the discharge.

The next step was the calculation of gas flows, their heating and cooling. The view of the three-dimensional calculation model in COMSOL Multiphysics is shown in Fig. 10, b. Carbon dioxide is fed into the plasmatron with a total flow of 2.6 l/min. Near the exit of the plasmatron, a heat release region is defined, the shape of which is determined by the zone of the electric field strength amplitude sufficient to sustain the discharge and by the plasma structure observed in the experiment. According to the distribution of streamlines (Fig. 11), the heat release region is enveloped by gas flows, due to which a large fraction of gas moves in the near-wall region and is less efficiently heated, which certainly reduces the efficiency of the heating process. After passing this region, the hot gas jet from the plasma torch interacts with four streams of quench gas (total flow rate of 3 l/min).

The results of numerical simulation of gas temperature and streamlines in the system are presented in Fig. 11. When quenching gas is supplied (11.b, 11.c), the volume of the high temperature region is significantly reduced compared to the temperature distribution without quenching (11.c), and the torch is effectively broken up and mixed with quenching gas and the surrounding medium. Comparison of the temperature distribution during quenching with nitrogen and carbon dioxide shows that in the second case the carbon dioxide quench jet shifts the high temperature region more strongly than nitrogen. This is due to the higher density of carbon dioxide, and consequently the larger momentum carried at the same flow of 3 l/min. Additionally, this effect is demonstrated in Figure 12, which shows the temperature vs. time dependence along the center streamline along the axis of the system. The cooling rates to 2000 K without quenching, with $N_2$ quenching and with $CO_2$ quenching along the streamline on the axis of the system were as follows, respectively, $0{,}56 \cdot 10^6\ K/s$, $1{,}26 \cdot 10^6\ K/s$ and $1{,}46 \cdot 10^6\ K/s$.

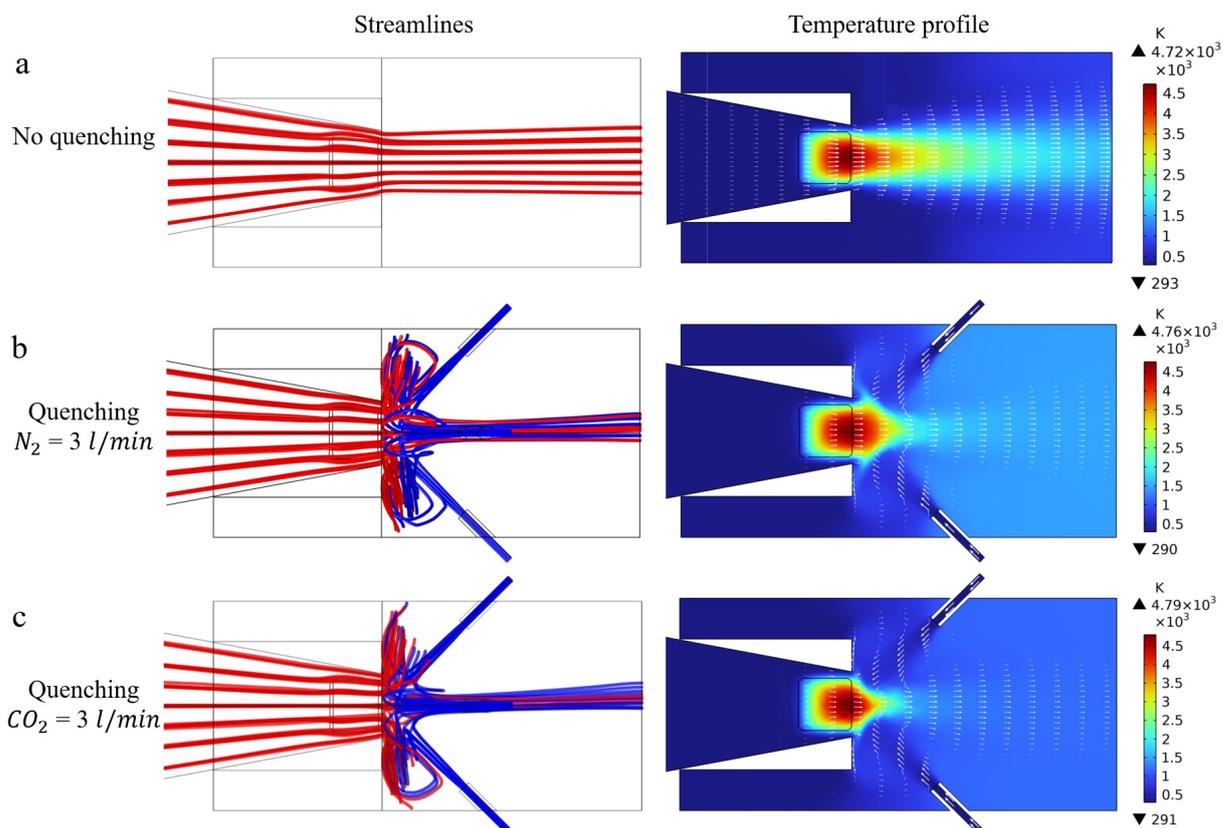

Figure 11. Results of numerical simulation of gas dynamic and heat exchange processes in COMSOL Multiphysics without quenching (a), quenching by nitrogen (b) and carbon dioxide flows (c). In the left part the image of streamlines from the source in the plasmatron - in red and from the quench gas inlet tube - in blue, in the right part the temperature distribution (indicated by color), as well as the velocity field in logarithmic scale (indicated by white arrows).

At the final stage of calculation, the time dependence of temperature along the streamlines uniformly distributed along the inlet cross-section of the plasma torch is calculated from the gas velocity data, and the chemical kinetics along the streamlines is calculated in the CHEMKIN package in the model of a homogeneous reactor. Pure $CO_2$ is set as the initial composition of the mixture, then, upon heating above 2000 K, $CO_2$ is converted to a mixture of $CO, O, O_2, CO_2$. The data on the final composition of the gas mixture at each streamline are used to calculate the degree of conversion using formula (1). The conversion rates obtained on all streamlines are averaged. The calculations show that only one third of the current lines reach a temperature sufficient to obtain a significant degree of conversion (>1%).

Fig. 13 shows an example of the calculated time dependence of molar fractions of components and temperature along the streamline, on which the maximum degree of conversion $K = 47.3\ \%$ was achieved during quenching with carbon dioxide. Two stages of chemical reactions are clearly distinguished: from 1 to 2.5 ms, direct reactions dominate, and almost complete decomposition of carbon dioxide occurs, while from 2.5 ms to 4 ms, the role of reverse reactions increases, recombination occurs, and the fraction of carbon dioxide rises to 43%. At temperatures below ~2000 K the composition of the mixture in the system on time scales of a few milliseconds practically does not change, so the key role in assessing the quenching efficiency is the cooling rate to this temperature. It can be seen that from a maximum temperature of 4150 K there was cooling to 2000 K at a rate of $\sim 1.6 \cdot 10^6\ K/s$, which is sufficient for effective quenching [27].

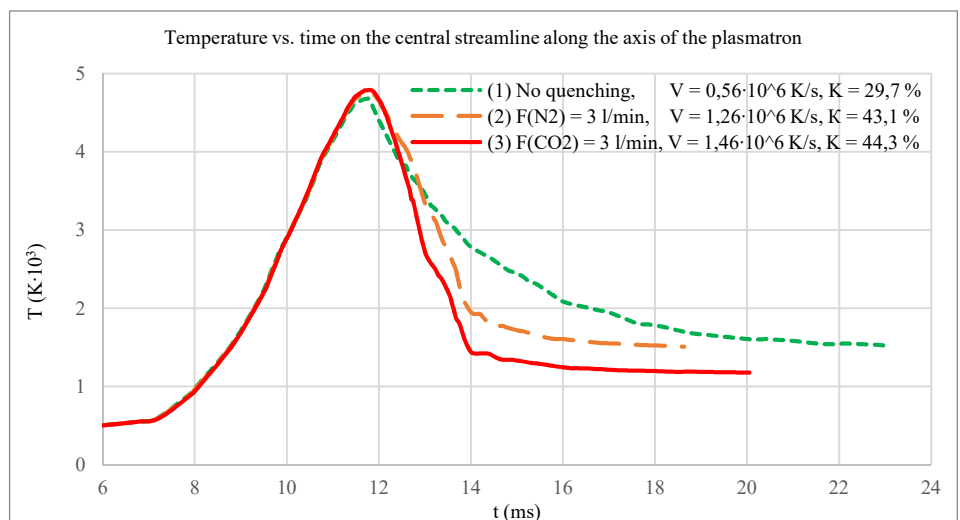

Figure 12. Temperature dependence along the system axis without quenching, with nitrogen and carbon dioxide quenching.

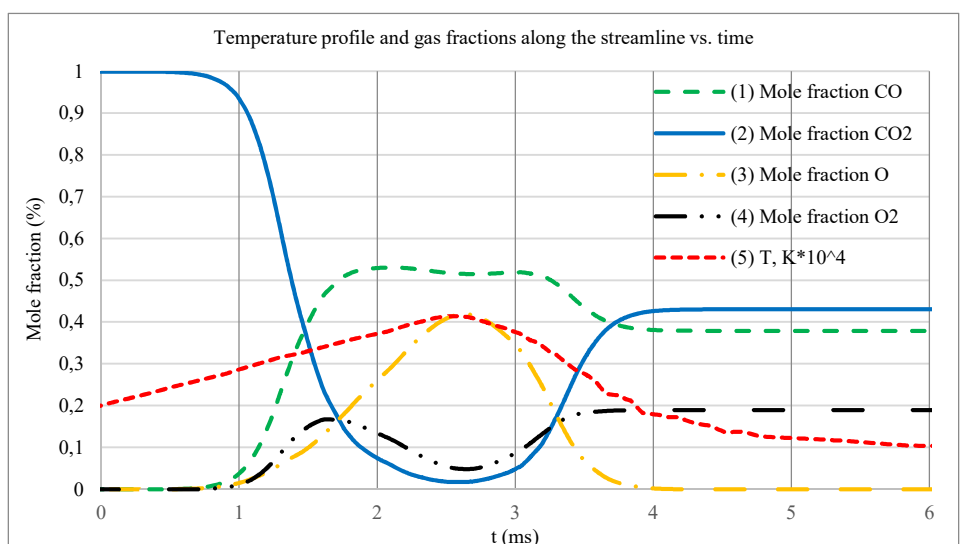

Figure. 13. Time dependence of molar fractions of components and temperature obtained in CHEMKIN for the streamline, where the highest degree of conversion at quenching with carbon dioxide is observed.

Figure 14 compares the degree of conversion obtained in the experiment with the conversion calculated from the results of numerical simulation (without quenching, with quenching with nitrogen 3 l/min, with quenching with carbon dioxide 3 l/min). Averaging over the set of streamlines was performed to calculate the integral value of the degree. It can be seen that the quenching gas supply leads to a multiple increase in the conversion degree according to both numerical simulation results and experiments. However, the calculated values of the degree of conversion differ significantly from those obtained in the experiment, and this difference is especially significant in the case of quenching with carbon dioxide.

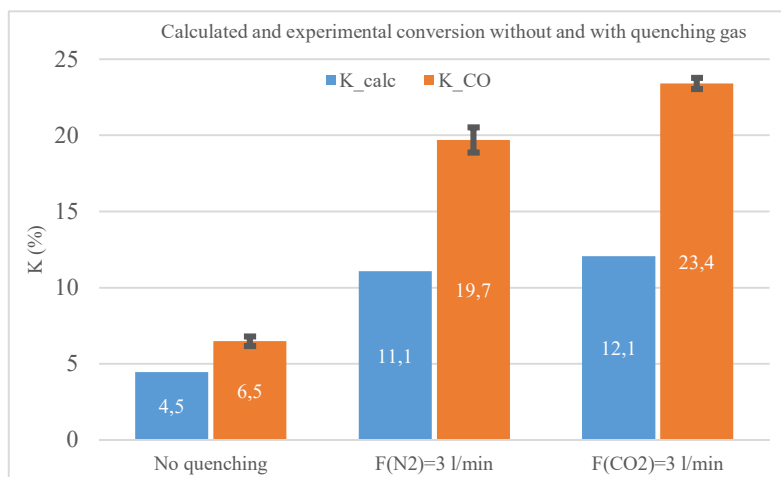

Figure. 14. $CO_2$ conversion calculated from numerical simulations of chemical kinetics in a 0D homogeneous reactor in CHEMKIN (K_calc) and experimental values of conversion rate ($K_{CO}$) at the same flows: $F_{CO_2} = 2.6\ l/min, F_{quenc} = 3\ l/min$.

Such a difference between the results of numerical modeling and the results of experiments can arise for a number of reasons, one of which is the inhomogeneous distribution of gas and plasma temperature in the discharge maintenance region. Thus, the shape of the heating region in the model has a great influence on the calculated values of the conversion rate (Fig. 10.b). In the model with a cylindrical heat source described in this paper, only about one third of the streamlines are heated to a temperature sufficient for a significant decomposition of carbon dioxide. This occurs, first, because the heat source does not occupy the entire cross-section of the waveguide, but only about 60%. And secondly, as a result of rapid heating of the gas, it is displaced from the heating zone into the near-wall region, which is evident from the envelope of the emission region by streamlines (Fig. 11). At the same time, the heating region used in the model may differ in size and shape from the real heat release region.

The numerical methods used in this work allow us to take into account only the reactions of thermal decomposition of carbon dioxide. At the same time, considering the nonequilibrium electronic component that is typical for the discharges supported by millimeter-wave radiation [22,23] can give an additional contribution to the vibrational excitation of $CO_2$ molecules and their subsequent dissociation. In addition, in a number of experiments, we have detected the ultraviolet radiation with wavelengths λ < 226 nm in the emission spectrum of the plasma inside the waveguide (a separate paper will be devoted to the study of the role of ultraviolet in such a discharge). This radiation can lead to the dissociation of molecular oxygen $O_2 + h\nu \rightarrow O + O$ [28], resulting in a possible reaction between carbon dioxide and atomic oxygen and the realization of an effective mechanism of "super ideal quenching". Such reactions can play an essential role both in the discharge itself and in the near-wall region of the plasmatron close to the discharge, where the gas does not get sufficient heating for thermal decomposition. This mechanism may play a key role in the difference between the experimental and calculated values of the conversion degree.

Finally, at high quenching gas flows, one cannot exclude its penetration not only into the post-discharge region, but also partially into the discharge inside the waveguide. In the case of quenching mixture ($CO_2$, $CO$, $O_2$), additional decomposition of $CO_2$ is possible, which in the experiment can lead to an increase in the degree of conversion. However, as the total $CO_2$ flux increases, the discharge becomes less stable, which explains its extinction at $F_{quenc} > 3\ l/min$. Also, a possible reason for the discharge extinction is the penetration of oxygen from the quenching mixture into the discharge region. At higher degrees of conversion, the oxygen concentration increases and electron losses for dissociative or three-particle attachment to $O_2$ molecules increase ($e + O_2 \rightarrow O^- + O, e + 2O_2 \rightarrow O_2^- + O_2$) [29], which leads to the changes in the ionization balance and discharge extinction.

## 7. CONCLUSIONS

A threefold increase in the $CO_2$ conversion and energy efficiency due to the cooling of the post-discharge region by the counterflow of gas has been demonstrated in the plasma of an atmospheric pressure discharge supported by microwave radiation of the gyrotron with the frequency of 24 GHz in the flow of carbon dioxide. The role of convective heat transfer in the process of gas mixture cooling in the post-discharge region has been experimentally demonstrated. At a nitrogen quench gas flow of 4.5 l/min, the $CO_2$ conversion rate was 23.8 %, energy efficiency was 19.7 %, and the formation of nitrogen oxides $N_xO_y$ was minimal. The possibility of using a stream of cooled gas mixture ($CO_2$, $CO$, $O_2$) taken from the expansion part of the reactor as a quenching gas was experimentally demonstrated, which made it possible to achieve a $CO_2$ conversion rate of 23.4 % and to eliminate the problem of dilution of reaction products by third-party gases. On the basis of numerical modeling, it is shown that the increase in the conversion degree upon the destruction of the plasma torch structure is due to the increase in heat exchange with the surrounding atmosphere, and the efficiency of this breakdown is determined by the velocity and density of quenching gas.

Further increase in the degree of conversion can be achieved by lowering the oxygen concentration [30], for example, by using different catalysts [31] or biochar [32,33] for oxygen binding. It is also possible to upgrade the quench gas supply system (angle of gas feed channels, distance to the plasmatron outlet, number of cooling channels) based on the results of numerical simulation of a correct heat transfer model that takes into account the thermal effect of reverse reactions and the real distribution of heat power density in the discharge.

Acknowledgements. This work was supported by Russian Science Foundation (project # 21-12-00376)